\documentclass[aps, 10pt, pra, twocolumn,preprintnumbers,amsmath,amssymb]{revtex4-1}
\usepackage{graphicx}
\usepackage{dcolumn}
\usepackage{bm}
\usepackage{color}
\usepackage{braket}
\usepackage{siunitx}
\usepackage[version=4]{mhchem}

\begin{document}

\title{Photon-momentum transfer in diatomic molecules: an {\it ab initio} study }

\author{Hao Liang$^{1}$}
\author{Mu-Xue Wang$^{1}$}
\author{Xiang-Ru Xiao$^{1}$}
\author{Qihuang Gong$^{1,2}$}

\author{Liang-You Peng$^{1,2}$}
\email{Email: liangyou.peng@pku.edu.cn }
\affiliation{$^{1}$State Key Laboratory for Mesoscopic Physics and Collaborative Innovation Center of Quantum Matter, School of Physics,  Peking University, Beijing 100871, China}
\affiliation{$^{2}$Collaborative Innovation Center of Extreme Optics, Shanxi University, Taiyuan, Shanxi 030006, China}

\date{\today}

\begin{abstract}

For a molecule, the two-center interference and the molecular scattering phase of the electron are important for almost all the processes that may occur in a laser field. In this study, we investigate their effects in the transfer of linear photon momentum to the ionized electron by absorbing a single photon.  The time-dependent Schr\"odinger equation of H$_2^+$ is numerically solved in {the multipolar} gauge in which the electric quadrupole term and  the magnetic dipole term are explicitly expressed. This allows us to separate the contributions of the two terms in the momentum transfer.  For different configurations of the molecular and the laser orientation,  the transferred momentum to the electron is evaluated at different internuclear distances with various photon energies { and two-center interferences are identified in the whole region. At small electron energies and small internuclear distances, we find significant deviations from the prediction of the classical double-slit model  due to the strong mediation of the Coulomb potential.} Finally, even for a large internuclear distance, our results show that a varying molecular scattering phase is important at all electron energies, which is beyond the simple prediction of the linear combination of the atomical orbitals.

\end{abstract}

\maketitle

\section{Introduction}

The dipole approximation is   widely used  while dealing with interaction between lasers and atoms or molecules in the non-relativistic region. However, it is invalid when the laser wavelength is comparable with  the size of atoms~\cite{seaton_momentum_1995,reiss_limits_2008, chelkowski_photon_2014, wang_nondipole_2018}, or when the laser intensity is increased to such an extent that the magnetic component of the Lorentz force can no longer be neglected~\cite{di_piazza_extremely_2012,reiss_tunnelling_2014}. Due to the progress of the  experimental technologies and methodologies,   the nondipole effects in atoms ionized by the infrared and mid-infrared lasers  have been observed recently~\cite{smeenk_partitioning_2011,ludwig_breakdown_2014}.
For a long time, the nondipole effects in photoelectron angular distributions from atoms and molecules have been investigated theoretically~\cite{cooper_multipole_1990,guillemin_nondipolar_2002,grum-grzhimailo_angular_2003,baltenkov_nondipole_2004,
seabra_molecular_2005,toffoli_photoelectron_2006,grum-grzhimailo_projection_2007} and experimentally~\cite{hemmers_large_2001,hosaka_nondipole_2006,zimmermann_dipole_2015}. By integrating over those asymmetric electron   distributions, one can get a net momentum of photoelectron along the direction of the laser propagation~\cite{seaton_momentum_1995,smeenk_partitioning_2011,ludwig_breakdown_2014}, which comes from the linear photon momentum partition between the photoelectron and the residual ion.

In order to study  the nondipole effect theoretically, different methods can be adopted, such as a numerical solution to the time-dependent Schr\"odinger equation~(TDSE) beyond the dipole approximation~\cite{forre_nondipole_2014,chelkowski_photon-momentum_2015,ivanov_nondipole_2016,wang_photon-momentum_2017}, or directly to the Dirac equation~\cite{ivanov_relativistic_2015}. In the tunnelling regime, one can also turn to  the adapted strong field approximation~\cite{walser_high_2000,kylstra_photon_2001,titi_quantum_2012,chelkowski_photon_2014,krajewska_radiation_2015,cricchio_momentum_2015,he_strong_2017} and the modified classical trajectory Monte Carlo method~\cite{liu_coulomb_2013,ludwig_breakdown_2014,tao_coulomb_2017}. In the high frequency regime, the nondipole effect can also be considered by the perturbation theory~\cite{seaton_momentum_1995, cooper_multipole_1990, guillemin_nondipolar_2002, chelkowski_photon_2014} or in the Kramers-Henneberger frame~\cite{henneberger_perturbation_1968,forre_exact_2005}.

The photon momentum transfer is one of the most important nondipole effects, which is directly related to the radiation pressure of light. However, most of previous work focused on the  momentum transfer in atomic systems.  Very recently, based on  an independent atom approximation,  Lao {\it et al.}~\cite{lao_longitudinal_2016} used the time-dependent perturbation theory to investigate  the interference effect in diatomic molecules to the  momentum transfer by absorbing one photon. For laser sources with wavelengths in the infrared regime,   Chelkowski and Bandrauk~\cite{chelkowski_photon-momentum_2018} studied the two-center effects and the dependence of  ellipticity  of the laser pulse,  by solving TDSE in the Cartesian coordinates with a soft-core Coulomb potential.

For diatomic molecules, the two-center nature can lead to   interferences in all the electron-related phenomena. With a high-energy photon incident, the ionization rate of the  electron shows an oscillation behavior with the variation of the photon energy~\cite{cohen_interference_1966,canton_direct_2011} and a double-slit like interference pattern in the photoelectron momentum spectrum can be seen~\cite{fojon_interferences_2006,fernandez_interferences_2007,hu_attosecond_2009,xu_attosecond_2011, yuan_linear-_2011,guan_double-slit_2012}. When the laser frequency is decreased and the intensity is increased, the interference effect  still exists in the multiphoton region~\cite{henkel_interference_2011} and even in tunneling region~\cite{bandrauk_lied:_2002,yurchenko_laser-induced_2004,bian_attosecond_2012}. It is possible to use such interferences to measure the internuclear distance~\cite{meckel_signatures_2014,wolter_ultrafast_2016}. Resonances due to the two-center interference were also shown to be present in the   Eisenbud-Wigner-Smith time delay of molecules~\cite{ning_attosecond_2014}.

In the present work, we present the first {\it ab initio} study  of the  photon-momentum transfer in the simplest benchmark system of H$_2^+$.  The corresponding 3-dimensional molecular TDSE  beyond the dipole approximation is exactly solved in the prolate spheroidal coordinates. For various orientations  of the molecular axis and the laser polarization,  the  momentum transfer to the electron is examined at different internuclear distances with various photon energies and two-center interference patterns are clearly observed. Different from previous studies, our results demonstrate that a varying molecular scattering phase is important at all electron energies, which is beyond the simple description of the linear combination of the atomical orbital. In addition, the present formulation of the interaction Hamiltonian is able to show that the contribution of the electric quadrupole term is dominant for all the cases considered in this work.
Atomic units (a.u.) are used throughout this paper unless otherwise stated.

\section{Theoretical methods}

In this section, we briefly describe our numerical methods. In particular, an appropriate gauge form to describe the laser-matter interaction is very important for the numerical convergence~\cite{cormier_optimal_1996} and to the physical interpretation~\cite{reiss_altered_2013}. This is also true when the nondipole terms are to be included in the Hamiltonian. In this work, we adopt such a gauge that the electric quadrupole term and  the magnetic dipole term is separate and  explicitly expressed. The resultant TDSE is solved at the fixed nuclear approximation in the prolate spheroidal coordinates using similar methods  to those previously used for the dipole case~\cite{telnov_ab_2007,tao_grid-based_2009,liang_accurate_2017}.

{One may wonder whether the momentum conservation between the laser field and the molecular system is violated under the fixed nuclear approximation. In fact, it has been shown for the atomic case that~\cite{chelkowski_photon_2014}, the fixed nuclear approximation is accurate with a relative error up to $O(\frac{m_e}{m_p})$ for the  photoelectron momentum distribution~(PMD). According to the momentum conservation, the linear momentum gained by the ion can be evaluated by the   difference between the  photon momentum and the linear momentum of the electron along the laser propagation.}

\subsection{Choice of the Gauge}
  In the radiation gauge, the time-dependent Hamiltonian of the electron interacting with a laser field  is given by
\begin{equation}
H = \frac{1}{2}[\bm p+\bm{A}(\bm{r},t)]^2  + V_0(\bm{r}),
\label{eq:tdse-nondiapole}
\end{equation}
where $\bm A$ is the vector potential of the laser field and $V_0(\bm{r})$ is the interaction potential between the electron and the nuclei.
Within the plane wave approximation of  the laser field, one can write the spatial and temporal distribution of the vector potential as
\begin{equation}
\bm{A}(\bm{r},t) = \bm{A}(\tau),
\end{equation}
in which $\tau = t-\alpha\hat {\bm{k}}\cdot\bm{r}$ is the retarded time with $\hat {\bm{k}}$ being the unit vector of  the laser propagation direction and $\alpha$ being the fine-structure constant  equal to the reciprocal of the speed of light in atomic units.

The vector potential can be expanded in terms  of $\alpha$ as
\begin{equation}
\begin{aligned}
\bm A(\tau) &=\bm A_0(t) + \bm r\cdot\nabla \bm A(\tau) + O(\alpha^2),\\
& = \bm A_0(t) + \alpha (\hat {\bm{k}}\cdot \bm r)\bm E_0(t) + O(\alpha^2),
\end{aligned}
\label{eq:series}
\end{equation}
where $\bm A_0(t), \bm E_0(t)$ respectively represents the vector potential and electric field at $\bm r=0$. Inserting Eq.~\eqref{eq:series} back into Eq.~\eqref{eq:tdse-nondiapole}, we can get the Hamiltonian:
\begin{equation}
\begin{aligned}
H =& \frac{\bm p^2}{2} + V_0(\bm{r}) +\bm A_0(t)\cdot\bm p + \frac{\bm A_0(t)^2}{2} \\
 &+\alpha(\hat {\bm{k}}\cdot \bm r)\left[\bm E_0(t)\cdot\bm p + \bm E_0(t)\cdot \bm A_0(t)\right]+O(\alpha^2).
\end{aligned}
\label{eq:gauge-usual}
\end{equation}
In principle,  the  Schr\"odinger equation is accurate up to  $O(\alpha)$, therefore,  there is no need to consider higher-order terms in the expansion~\eqref{eq:series}.

The Hamiltonian~\eqref{eq:gauge-usual} is widely used when  one considers the nondipole effect in the ionization dynamics~\cite{seaton_momentum_1995,forre_nondipole_2014,chelkowski_photon-momentum_2015,cricchio_momentum_2015,ivanov_nondipole_2016,wang_photon-momentum_2017}.  In the literature, there are many other choices of gauge~\cite{forre_exact_2005,selsto_alternative_2007,forre_generalized_2016}, though it is hard to relate a physical meaning  to  each of those nondipole terms.
{Here, we consider another gauge form.} Under a gauge transformation
\begin{equation}
\Psi' = \mathrm e^{\mathrm i\Lambda}\Psi,
\label{transform}
\end{equation}
where $\Lambda$ is an arbitrary real function of the space and time, the Hamiltonian is correspondingly  transformed to
\begin{equation}
\begin{aligned}
H' &= \mathrm e^{\mathrm i\Lambda}(H-\mathrm i\partial_t)\mathrm e^{-\mathrm i\Lambda},\\
&=\mathrm e^{\mathrm i\Lambda}H\mathrm e^{-\mathrm i\Lambda}-\partial_t\Lambda.
\end{aligned}
\end{equation}

{If one chooses $\Lambda =\int_0^1\bm A(\eta\bm r,t)\cdot\bm r\,\mathrm d\eta$, i.e. the Powers-Zienau-Wolley gauge transformation~\cite{power_coulomb_1959,woolley_molecular_1971},} and applies it to Eq.~\eqref{eq:tdse-nondiapole}, after some simple derivations, one can get a new Hamiltonian
\begin{equation}
\begin{aligned}
H'&=\underbrace{\frac{\bm p^2}{2}+V_0(\bm{r})}_{H_0}+\underbrace{\bm r\cdot\bm E_0(t)}_{\text{E-dipole}} \underbrace{-\frac{1}{2}\alpha(\hat {\bm{k}}\cdot\bm r)[\bm r\cdot\partial_t\bm E_0(t)]}_\text{E-quadrupole}\\
&+\underbrace{\frac{1}{2}\alpha\bm L\cdot[\hat {\bm{k}}\times\bm E_0(t)]}_\text{B-dipole}+O(\alpha^2),
\end{aligned}
\label{eq:hami-final}
\end{equation}
in which $\bm L=\bm r\times\bm p$ is the angular momentum operator.  {This is so called multipolar gauge.}  In this formulation,  it is clear that one can separate the terms of the electric dipole, the electric quadrupole, and the magnetic  dipole. As will be seen later, it is convenient to investigate separately how the inhomogeneity of electric field and magnetic field will affect the the electron dynamics.

For the single-photon ionization process with a photon energy $\omega$, some exploitations of symmetry can be made by the lowest-order perturbation theory~(LOPT) before solving the TDSE. It is easy to show that the photoelectron momentum distribution is given by
\begin{equation}
\begin{aligned}
P(\bm q) = &\delta(\omega+E_i-\bm q^2/2)\times\\
&\left|\Braket{\bm q|\bm r +\mathrm i\omega\frac{\alpha}{2}(\hat {\bm{k}}\cdot\bm r)\bm r+\frac{\alpha}{2}\bm L\times\hat {\bm{k}}|i}\cdot\tilde{\bm E_0}\right|^2,
\end{aligned}
\end{equation}
where $\ket{\bm q}$ and $\ket{i}$ respectively represents a continuum state with a momentum $\bm q$ and the initial bound state, and $\tilde {\bm E_0}$ is the complex amplitude of the electric field. Then  the average momentum of the photoelectron can be evaluated as
\begin{equation}
\braket{\bm q} = \frac{1}{\sigma_1} \int \bm qP(\bm q)\,\mathrm d^3\bm q,
\end{equation}
where $\sigma_1=\int P(\bm q)\,\mathrm d^3\bm q$ is the total one-photon ionization cross section. For a system with a space inversion symmetry, only the overlapping terms between the dipole term and the nondipole terms in the numerator survive under the integration, therefore
\begin{equation}
\begin{aligned}
\braket{\bm q} =& \frac{\alpha}{\sigma_1}\mathrm{Re}\int\mathrm d^3\bm q\,\delta(\omega+E_i-\bm q^2/2)\times\\
&\Braket{i|\bm r\cdot\tilde{\bm E_0^*}|\bm q}\bm q\Braket{\bm q|\mathrm i\omega(\hat {\bm{k}}\cdot\bm r)\bm r+\bm L\times\hat {\bm{k}}|i}\cdot\tilde{\bm E_0}\\
=& \braket {\bm q}_\text{E4}+\braket {\bm q}_\text{M2},
\end{aligned}
\label{eq:perturbation}
\end{equation}
in which the contribution from electric quadrapole and magnetic dipole terms can be simply added together. For the electron with an energy $E$ ionized from 1s state of hydrogen-like atoms, it has been shown that $\braket {\bm q}_\text{E4} = 1.6\alpha E\hat{\bm k}$~\cite{chelkowski_photon_2014}. However, one has  $\braket {\bm q}_\text{M2}=\bm 0$ due  to zero angular momentum component along the direction of the magnetic field $\hat {\bm{k}}\times\bm E$. In this case the electron gains a momentum from the inhomogeneous electric field instead of the magnetic field. Things would be much more complex for other systems.  The average momentum of the  ionized electron is not necessarily collinear with the photon momentum, since the interaction between the  electron and the residual ions can be anisotropic.

\subsection{Computational Details}

After choosing the {multipolar gauge}, we numerically solve the TDSE with the Hamiltonian given by Eq.~\eqref{eq:hami-final} for  H$_2^+$ with a fixed-nuclear approximation in  the prolate spheroidal coordinates$(\xi,\eta,\phi)$:
\begin{equation}
\xi=(r_1+r_2)/R,\quad \eta=(r_1-r_2)/R,
\end{equation}
where $r_i~(i=1,2)$ is the distance between the electron and the $i$-th nucleus, $R$ is the internuclear distance, and $\phi$ is the azimuthal angle around the molecular axis.  In the Cartesian coordinates  of the molecular frame, we assume that the molecular axis  is collinear with $z$-axis.

The time dependent wavefunction is expanded in a production  basis of the finite element discrete variable representation~(FE-DVR)~\cite{rescigno_numerical_2000} and the spherical harmonics $Y_l^m$, i.e.
\begin{equation}
\begin{aligned}
&f_{I,l,m}(\xi,\eta,\phi)= \\
&\begin{cases}
\chi_I(\xi)Y_l^m(\arccos\eta,\phi),& \text{for even }m; \\
\sqrt{\frac{\xi-1}{\xi+1}}\chi_I(\xi)Y_l^m(\arccos\eta,\phi),& \text{for odd }m,
\end{cases}
\end{aligned}
\end{equation}
where the multiplication factor $\sqrt{(\xi-1)/(\xi+1)}$ is used to remove the singularity near the nuclei~\cite{telnov_ab_2007,tao_grid-based_2009} and $\chi_I(\xi)$ is the FE-DVR basis function. For a full convergence, the  angular basis is typically taken to be $l_{\max}=60$ and $m_{\max}=2$ in the present calculations.

To evaluate the magnetic dipole term in Eq.~\eqref{eq:hami-final}, one has to   express the angular momentum operator in the prolate spheroidal coordinates.  The $z$-component is straightforward, i.e.,
\begin{equation}
L_z = -\mathrm i\partial_\phi.
\end{equation}
To compute the $x,y$ components, we introduce the ladder operators for simplicity
\begin{equation}
\begin{aligned}
L_\pm &\equiv L_x\pm\mathrm iL_y,\\
&=\frac{1}{\xi^2-\eta^2}\Big\{\pm\mathrm e^{\pm\mathrm i\phi}\big[\eta\sqrt{1-\eta^2}(\xi^2-1)^{1/4}\partial_\xi(\xi^2-1)^{1/4}\\
&-\xi\sqrt{\xi^2-1}(1-\eta^2)^{1/4}\partial_\eta(1-\eta^2)^{1/4}\big]\\
&+\xi\eta\left(\sqrt{\frac{1-\eta^2}{\xi^2-1}}+\sqrt{\frac{\xi^2-1}{1-\eta^2}}\right)\mathrm e^{\pm i\phi/2}\mathrm i\partial_\phi\mathrm e^{\pm\mathrm i\phi/2}\Big\}.
\end{aligned}
\end{equation}

Once one gets the final-state wavefunction at the end of the laser pulse, the PMD is obtained by projecting it onto the molecular scattering states. The details of numerical methods can be found in our  previous work~\cite{liang_accurate_2017,hou_attosecond_2012,yan_grid_2009}.

\section{Results and Discussions}

\begin{figure}[htbp]
\includegraphics[width=\linewidth]{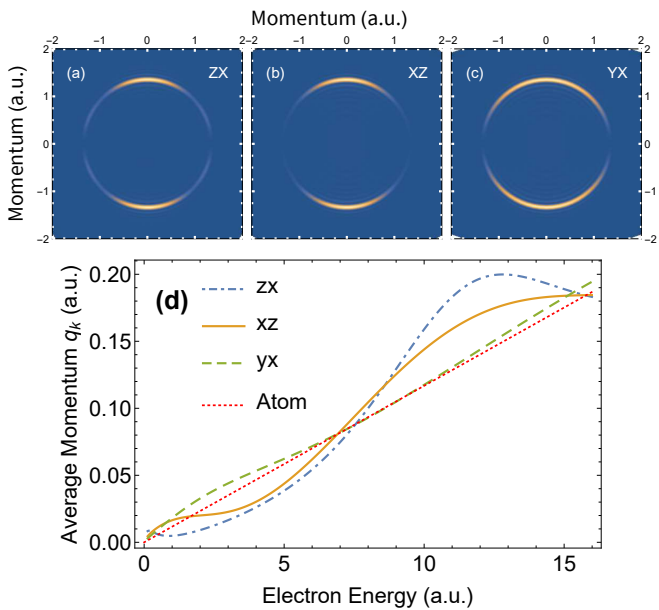}
\caption{
The 2D photoelectron momentum distributions in the polarization-propagation plane of  H$_2^+$ by a linearly polarized laser in: (a) the $zx$, (b) the $xz$, and (c) the $yx$ configuration respectively. The laser is polarized along the vertical axis and the wavevector is along the horizontal axis. (d) The average momentum transfer of the photoelectron along the laser propagation direction~($q_k$) as a function of the electron energy, { ionized from H$_2^+$ $1s\sigma_g$ state for the three typical configurations: the $zx$ (dot-dashed blue line), the $xz$ (solid yellow line), and the $yx$ (dashed green line)} respectively. The result for H atom~(red dotted line) is also plotted for comparison.}
\label{fig1}
\end{figure}

 In this section, we will present our main results about the photon momentum transfer in the simplest diatomic molecule H$_2^+$. As there involves three important physical vectors, i.e., the molecular axis $\hat {\bm{R}}$, the laser wavevector $\hat {\bm{k}}$,  and the laser polarization direction $\hat {\bm{E}}$, thus  there will be various configurations of the laser-molecule interacting system. Without loss of generality, we only consider three typical configurations: (1) $\hat {\bm{k}}  {\it \parallel} \hat {\bm{R}}$, labeled by the $xz$ configuration;  (2) $\hat {\bm{E}}  {\it \parallel} \hat {\bm{R}}$, labeled by the $zx$ configuration;  (3) $\hat {\bm{k}}  {\it \perp} \hat {\bm{R}}$  and $\hat {\bm{E}}  {\it \perp} \hat {\bm{R}}$, labeled by the $yx$ configuration.

 For all the following calculations, we use a $20$-cycle laser pulse  with a Gaussian envelope at a relative weak intensity $I=\SI{5e12}{\watt\per\square\centi\meter}$. The average momentum $\braket{\bm q}$ of the ionized electron is evaluated at different internuclear distances with a tunable laser frequency $\omega$. We choose its component along the laser propagation direction $\hat {\bm k}$ for analysis:
\begin{equation}
q_k\equiv\braket{\bm q}\cdot\hat {\bm{k}}.
\end{equation}
Although those momenta are indeed parallel to laser propagation direction in these three configurations, the momentum transfer { for the one-photon case } in other configurations can be evaluated by a linear combination of these three configurations according to Eq.~\eqref{eq:perturbation}.

Let us first look at the cases where $R=R_e= 2$~a.u. with the initial state $1s\sigma_g$ for all the three configurations defined above.  For a fixed photon energy of $\omega=2 $~a.u., the 2-dimensional PMD is respectively shown in Figs.~\ref{fig1}(a)-(c) for the three cases, in which electrons are all ejected mainly along the laser polarization axis.  In fact, the ionization probability has a tiny  preference in the direction of the laser  propagation direction, which is  invisible to the naked eye. In order to quantify this preference,  by integrating over the PMD,  one can extract the average momentum $\braket{\bm q}$, whose component $q_k$ along the laser propagation can be evaluated. One can calculate $q_k$ as a function of  the energy of the electron ionized by a photon with a various  energy $\omega$. It will be instructive to compare the $q_k(E)$ for all the three configurations against that of the atomic case, which are shown  in Fig.~\ref{fig1}(d).

From Fig.~\ref{fig1}(d),  we can see that the results for all the three configurations are oscillating around that of the atomic case $q_a=1.6\alpha E$~\cite{chelkowski_photon_2014}.  Apparently, the oscillation  is mainly due to the two-center interference effect of the molecule, which has been discussed recently using an independent atom model~\cite{lao_longitudinal_2016}. Approximately,  there will be a phase difference  of $qR\cos\theta_e$ between photoelectrons with a momentum $\bm q$ ionized from  the two different nuclei, where $\theta_e$ is the angle between the  photoelectron momentum and molecular axis.  This is the reason why the amplitude for the $zx$ configuration~($\hat {\bm{E}}  {\it \parallel} \hat {\bm{R}}$) is the largest, as $|\cos\theta|\approx 1$ for most of electrons.

\begin{figure}[htbp]
\includegraphics[width=\linewidth]{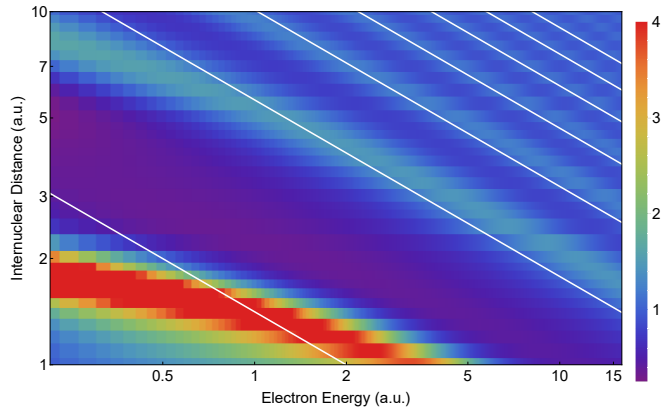}
\caption{The ratio between the average momentum transfer~($q_k$)  from H$_2^+$ to  that  from H atom in the $(E,R)$ plane, plotted in a  double logarithmic scale. White straight lines with a slope equal to $-1/2$ are plotted for reference.}
\label{fig2}
\end{figure}

To investigate the oscillatory phenomenon in details, we introduce a ratio $F$ to describe the difference between the results of the molecule and the atom,
\begin{equation}
F \equiv   \frac{q_k}{q_a} =  \frac{5q_k}{8\alpha E}.
\end{equation}
For the case of the $zx$ configuration,    we plot in Fig.~\ref{fig2} the ratio $F$ as a function of the internuclear distance $R$ and the electron energy $E$ in the double logarithm scale.  Clear stripes with a slope approximate to $-1/2$ can be observed. This means that the electrons emitted from different nuclei interfere  with a constructive or a destructive phase, i.e.,
\begin{equation}
qR\cos\theta_e =  n \pi, \Longrightarrow   \sqrt{2E}R=\text{const.}
\end{equation}
  However, it is important to note that  the stripes are bending away from the simple prediction  in the range of small internuclear distances or small electron energies, where   the Coulomb potential is inevitable to change the emission path and phase of the electrons.

\begin{figure}[htbp]
\centering
\includegraphics[width=.8\linewidth]{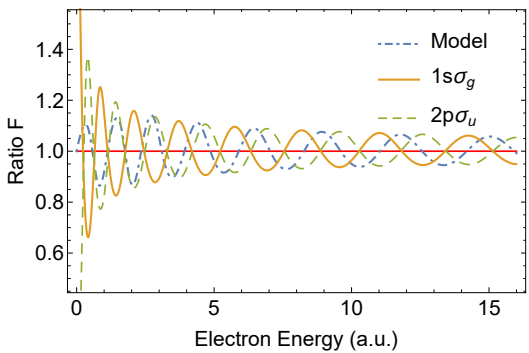}
\caption{ The ratio between the average momentum transfer  for  $1s\sigma_g$~(yellow solid line) and $2p\sigma_u$~(green dashed line) of H$_2^+$ with internuclear distance $R=10$~a.u.,  calculated by TDSE.  The prediction by the independent atom model for $1s\sigma_g$~(blue dot-dashed line)  is also shown, together with a  horizontal line with $F=1$  to guide one's eyes.}
\label{fig3}
\end{figure}

  Actually, even in the rest of regions of large $R$ and $E$,  the scattering phase of the long-range Coulomb potential  also plays its important role.  To demonstrate this, we show  our {\it ab initio} results of $F$  for both $1s\sigma_g$ and  $2p\sigma_u$  at $R=10 $~a.u. in Fig.~\ref{fig3}. Interestingly, an exact $\pi$ phase difference  between the two results is observed, which comes from the phase difference between the two initial states since, in the limit of a large $R$, these two states can be expressed as
\begin{equation}
\ket{1s\sigma_g}  = (\ket{1}+\ket{2})/\sqrt{2},\quad \ket{2p\sigma_u}  = (\ket{1}-\ket{2})/\sqrt{2},
\end{equation}
where $\ket{1}$ and $\ket{2}$  represent the 1s state of hydrogen atom for the $1$st and $2$nd nucleus respectively.

What is also shown in Fig.~\ref{fig3}  is the prediction of  the independent atom model~\cite{lao_longitudinal_2016} for the $1s\sigma_g$. For the independent atom approximation, the final PMD is simply obtained by coherently adding PMDs from two hydrogen atoms  with a phase factor $\mathrm e^{\mathrm iqR\cos\theta_e}$.  By comparing results from the TDSE  and the independent atom model, one can observe a phase  difference varying from $\pi$ to $\pi/2$ as the electron energy is increased. This means that molecular scattering phase is important at all electron energies, which is beyond the simple prediction of the linear combination of the atomical orbital.   In fact, a similar phase difference can also be observed in the dipole ionization rate~\cite{fojon_interferences_2006} and in the photoelectron angular distribution~\cite{yuan_linear-_2011,guan_double-slit_2012}.  It can be qualitatively explained as the scattering phase for the electron that is ionized from the back nucleus and subsequently scattered by the forward nucleus~\cite{henkel_interference_2011}.

\begin{figure}[htbp]
\centering
\includegraphics[width=.8\linewidth]{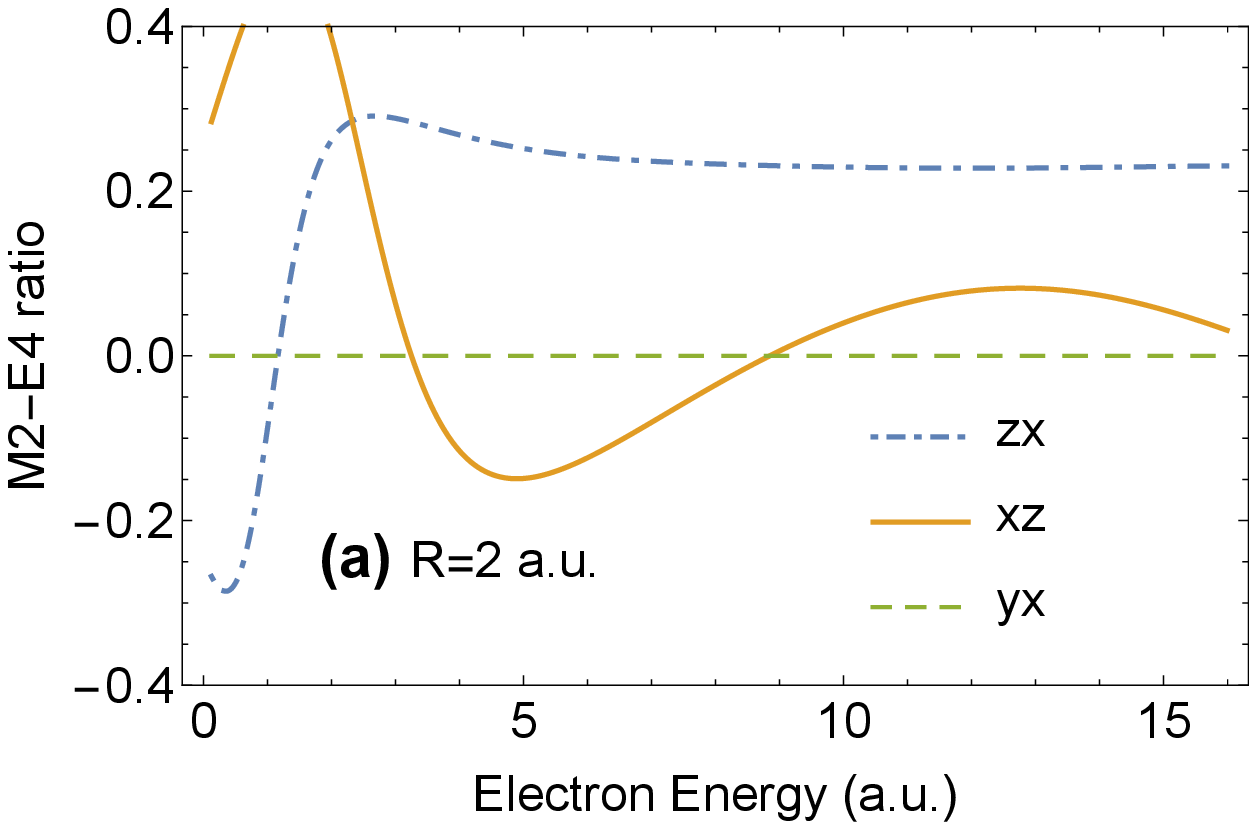}
\includegraphics[width=.8\linewidth]{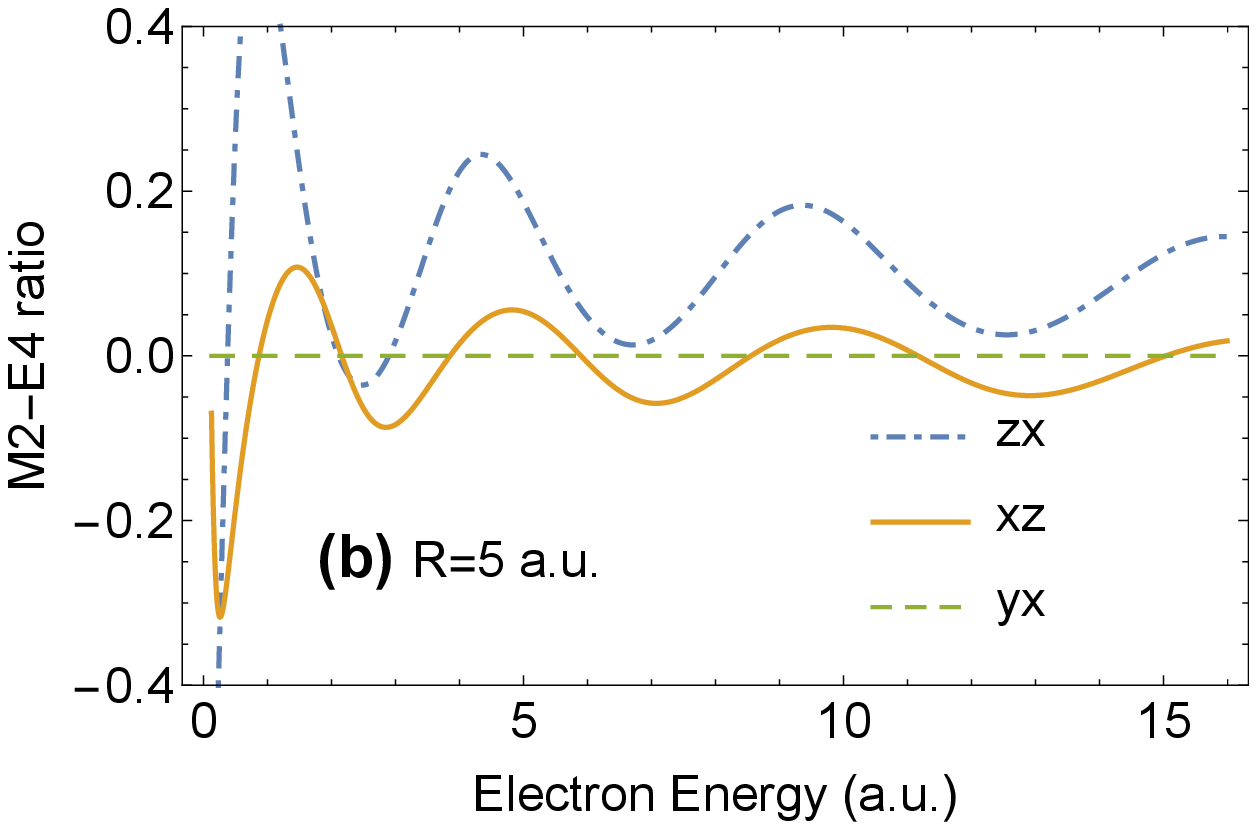}
\includegraphics[width=.8\linewidth]{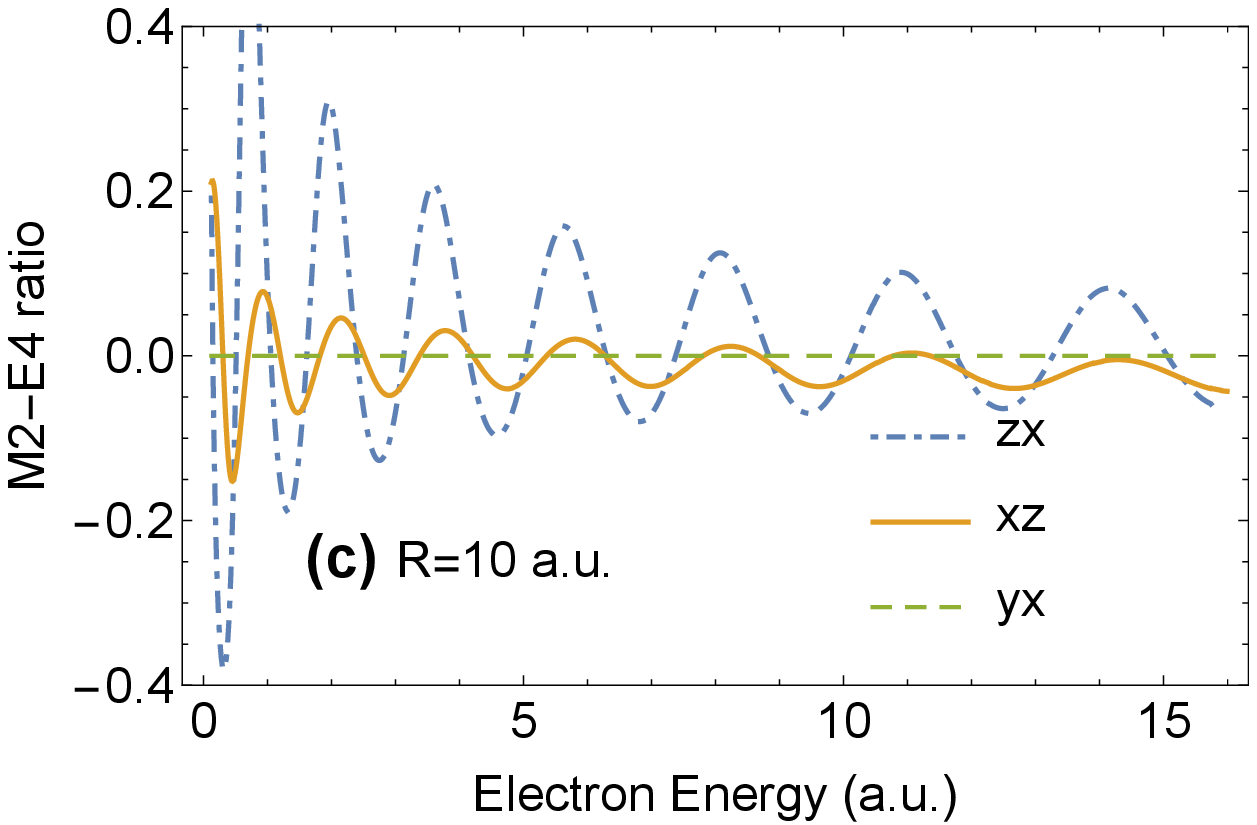}
\caption{The ratio between the contribution by electric-quadrupole term to the average momentum transfer and that by the magnetic dipole term, with different internuclear distances: (a) $2\,\rm{a.u.}$, (b) $5\,\rm{a.u.}$, (c) $10\,\rm{a.u.}$. For each frame, results are respectively presented  for the { $zx$~(dot-dashed blue line), the $xz$~(solid yellow line), and the $yx$~(dashed green line)} configuration.}
\label{fig4}
\end{figure}

Now, we can take advantage of the present formulation for the interaction Hamiltonian to examine the different contributions to the linear momentum transfer by the  electric quadrupole term   and  the magnetic dipole term. According to  Eq.~\eqref{eq:perturbation}, the average momentum shift $q_k$ of the photoelectron equals to the sum of electric quadrupole~(E4) contribution  and the magnetic dipole~(M2) contribution. By turning off one of these two terms in the Hamiltonian, we can evaluate $q_k$  separately by solving the corresponding TDSE. In Figs.~\ref{fig4}(a)-(c), for the $1s\sigma_g$,  we plot the ratio between the contribution of M2 and E4 to the average momentum shift at different $R$ and various electron energies.  For all the three configurations, the electric quadrupole term contributes to the average momentum shift dominantly, while the magnetic dipole term only contributes to the oscillation amplitude  about $20\%$.  It should be noted that for the $yx$ configuration where the magnetic field is parallel to the molecular axis, the magnetic field does not contribute at all since the angular momentum along the $z$ axis of the $\sigma$ state is zero.

 {Finally, let us turn to the practical issues that may be faced for an experimental observation.  Nowadays, the free-electron lasers~(FEL)~\cite{prince_coherent_2016} can provide the coherent xuv sources with durations under a few femtoseconds.   The photon ionization process of the electron,  is usually much faster than the vibration period of the ground vibrational state of H$_2^+$. For all the cases considered in the present work,  we use  20-cycle xuv pulses with various photon energies. It is easy to estimate that the pulse duration varies from 0.2~fs to  1.6~fs, which is at most 1/10 of the vibration period of the ground state  H$_2^+$.  Thus, it is reasonable to assume a vertical Franck-Condon transition,  in which case the two nuclei are almost fixed during the ionization process of the electron. }

 {Nevertheless, it will be instructive to approximately account for the influence of  the ground vibrational state to the oscillation feature of the transferred  linear momentum of the electron, by  averaging over a Gaussian-like vibrational wave function. To be specific, for various  $\omega$, one can evaluate
\begin{equation}
\braket{q_k(\omega)}  = \frac{\int q_k(R, \omega) W(R) \sigma (R, \omega)\,\mathrm d R}{\int W(R) \sigma (R, \omega)\,\mathrm d R }, \label{avg}
\end{equation}
 where $W(R)$ is the probability density according to  the Gaussian  nuclear wave function around $R_e = 2$ a.u. and $\sigma (R, \omega)$ is the total single-photon ionization cross section at $R$ for a photon energy $\omega$.   After taking the quantum nature of the ground vibrational state, we find that the oscillation feature of the electron momentum along the laser propagation is still clearly present, with only very small changes in the oscillatory amplitude and in the crossing points with the curve for the atomic case. }

  {For the interferences  at large $R$ other than $R_e$, it is also feasible to be measured by a pump-probe scheme.  A ultraviolet pulse can be used to pump the electron from the $1s\sigma_g$ state onto the dissociative $2p\sigma_u$ state, which be subsequently ionized  after some delay by  a probe pulse at larger $R$. As shown by Fig.~\ref{fig3}, the oscillation of the linear momentum of electron ionized from $2p\sigma_u$ will be out of phase with that from $1s\sigma_g$ state. }

  {In practice, one can measure the differential momentum distributions using the apparatus such as the cold target recoil ion momentum~(COLTRIM)~\cite{dorner_cold_2000} spectroscopy or the  velocity map imaging~(VMI)~\cite{eppink_velocity_1997}, from which the linear momentum transfer can be extracted from the electron momentum distribution along the laser polarization.  To observe the interference structure, one needs to sweep the photon energy, which can be easily achieved in synchrotron light sources or FEL. We thus expect that our theoretical observations  can be potentially observed with current experimental techniques.
}

\section{Conclusion}

In summary, by using a gauge form  in which the electric quadrupole term and  the magnetic dipole term can be treated separately, we carried out an {\it ab initio} study on the linear photon momentum transfer in the simplest diatomic molecule H$_2^+$ {in the single photon ionization process. The present theoretical methods can be extended to  investigate the nondipole effects in the multiphoton and tunneling regime.}  Different from the atomic case, the transferred momentum to the electron shows an oscillatory structure as a function of the internuclear distance and the electron energy, originating from the two-center interference of the diatomic molecule. From our exact results, we demonstrate  significant deviations from the prediction of the classical double-slit model  due to the strong mediation of the Coulomb potential. Even for a large internuclear distance, our results show that the molecular scattering phase is crucial at all electron energies, which is beyond the simple prediction of the linear combination of the atomical orbitals. Finally, the present formulation of the nondipole correction to the Hamiltonian allows us to identify the dominant contribution by the  electric quadrupole term to the linear photon momentum transfer for the molecule. { We point out that, with the currently available technologies, the present theoretical predictions may be confirmed experimentally in the near future.}

\begin{acknowledgments}

This work is supported by the National Key R\&D Program of China~(Grant No. 2018YFA0306302), and by the National Natural Science Foundation of China~(NSFC) under Grant Nos. 11725416, and 11574010.  L.Y.P. acknowledges the support by the National Science Fund for Distinguished Young Scholars.

\end{acknowledgments}

\bibliography{ref}

\end{document}